\documentclass[aps,prl,reprint,superscriptaddress,nofootinbib]{revtex4-2}
\usepackage{graphicx,tikz}
\usepackage[T1]{fontenc}
\usepackage{lmodern}
\usepackage{amsmath,amssymb,bm,booktabs,microtype}
\usepackage[colorlinks=true,linkcolor=blue,citecolor=blue,urlcolor=blue]{hyperref}
\hypersetup{pdftitle={Cylinder Defect Casimir Energy and Weyl Anomalies},pdfauthor={Yang Zhou}}
\usetikzlibrary{arrows.meta}
\definecolor{ink}{HTML}{213547}
\definecolor{ambient}{HTML}{688BA5}
\definecolor{defect}{HTML}{B96015}
\newcommand{\dd}{\mathrm{d}}
\newcommand{\R}{\mathbb{R}}
\newcommand{\cD}{\mathcal{D}}
\newcommand{\cT}{\mathcal{T}}
\newcommand{\Ric}{\mathcal{R}}

\newcommand{\Ksq}{\mathring{\Pi}^{\,2}}
\newcommand{\Wpull}{W_{ab}{}^{ab}}
\newcommand{\arxiv}[1]{\href{https://arxiv.org/abs/#1}{arXiv:#1}}
\begin{document}
\title{Cylinder Defect Casimir Energy and Weyl Anomalies}
\author{Yang Zhou}
\email{yang\_zhou@fudan.edu.cn}
\affiliation{Department of Physics, Fudan University, Shanghai 200433, China}

\begin{abstract}
We distinguish two Casimir observables associated with a conformal surface defect. A circular cylinder embedded in flat space has a logarithmic contribution to its Casimir energy, $-d_1\log(R/\epsilon)/(24R)$, fixed by the displacement anomaly. A great cylinder obtained by radial quantization of the ambient theory instead has Casimir energy $-b/(12R)$ in the conformal defect vacuum with a fixed local anomaly convention. Free-field and holographic calculations provide checks of the proposed relations in both embeddings. 
\end{abstract}
\maketitle

\textit{Introduction.---}The ground-state energy of a two-dimensional
conformal field theory (CFT) on a circle of radius $R$ is
$-c/(12R)$: a local Weyl anomaly thus determines a global finite-size
observable~\cite{Blote,Affleck}. Whether this relation extends to a
two-dimensional conformal defect coupled to a higher-dimensional CFT
is less clear. Energy can flow between the defect and the ambient
degrees of freedom, so the defect need not possess a separately
conserved stress tensor or exhibit Virasoro
symmetry~\cite{Billo,JensenEntropy}. Moreover, its Weyl anomaly depends
on both the induced metric and the
embedding~\cite{GrahamWitten,Henningson,Schwimmer}.
These features are central to the study of surface critical phenomena,
Wilson surfaces, and replica
defects~\cite{Metlitski,GiombiLiu,Trepanier,Raviv-Moshe:2023yvq,BianchiRenyi,DrukkerTech}.

\begin{figure*}[t]
\centering
\definecolor{cylBlue}{HTML}{1677E8}
\definecolor{cylRed}{HTML}{EF4B2D}
\definecolor{cylInk}{HTML}{202124}
\resizebox{\textwidth}{!}{%
\begin{tikzpicture}[x=1cm,y=1cm,font=\small,text=cylInk,
  line cap=round,line join=round,>={Latex[length=1.7mm]},
  ambient/.style={draw=cylBlue,line width=.9pt},
  defect/.style={draw=cylRed,line width=1.2pt},
  hidden/.style={dash pattern=on 2pt off 2pt,line width=.75pt}]
\path[use as bounding box] (0,0) rectangle (15.1,5.95);

\foreach \dx in {0,7.7} {
  \begin{scope}[shift={(\dx,0)}]
    \draw[->,line width=.75pt] (.70,1.02)--(.70,4.95)
      node[above=2pt] {$\tau$};
  \end{scope}
}

\begin{scope}
  \node[font=\normalsize] at (3.65,5.65)
    {\textbf{(a)} Bent cylinder};
  \node at (3.65,5.23) {$\Sigma\subset\mathbb R^d$};

  \path[draw=cylBlue!75,fill=cylBlue!4,line width=.7pt]
    (1.05,.90)--(5.05,.90)--(6.25,2.20)--(2.25,2.20)--cycle;
  \path[fill=cylRed!7]
    (2.35,1.55)--(2.35,4.05)
    arc[start angle=180,end angle=0,x radius=1.30,y radius=.43]
    --(4.95,1.55)
    arc[start angle=0,end angle=-180,x radius=1.30,y radius=.43]
    --cycle;
  \draw[defect] (2.35,1.55)--(2.35,4.05);
  \draw[defect] (4.95,1.55)--(4.95,4.05);
  \foreach \h in {1.55,4.05} {
    \draw[defect,hidden] (4.95,\h)
      arc[start angle=0,end angle=180,x radius=1.30,y radius=.43];
    \draw[defect] (2.35,\h)
      arc[start angle=180,end angle=360,x radius=1.30,y radius=.43];
  }
  \node[text=cylRed,font=\large] at (3.65,2.95) {$\Sigma$};
  \fill[cylInk] (3.65,1.55) circle[radius=.85pt];
  \draw[->,line width=.6pt] (3.65,1.55)--(4.93,1.55)
    node[pos=.43,above=1pt,inner sep=1pt] {$R$};
  \draw[->,draw=cylRed,line width=.9pt] (4.95,1.55)--(5.76,1.55)
    node[right=2pt,text=cylRed] {$n$};
  \node at (3.65,.28)
    {$\mathcal R_\Sigma=0,\qquad
      \mathring\Pi^{\,2}=\dfrac{1}{2R^2}$};
\end{scope}

\begin{scope}[shift={(7.7,0)}]
  \node[font=\normalsize] at (3.65,5.65)
    {\textbf{(b)} Great cylinder};
  \node at (3.65,5.23)
    {$\Sigma\subset\mathbb R_\tau\times S_R^{d-1}$};
  \draw[defect,hidden] (2.73,1.55)--(2.73,4.05);
  \draw[defect,hidden] (4.57,1.55)--(4.57,4.05);
  \foreach \h in {1.55,4.05} {
    \path[ambient,fill=cylBlue!4] (3.65,\h) circle[radius=.92];
    \draw[defect,hidden] (4.57,\h)
      arc[start angle=0,end angle=180,x radius=.92,y radius=.30];
    \draw[defect] (2.73,\h)
      arc[start angle=180,end angle=360,x radius=.92,y radius=.30];
  }
  \node[text=cylRed,font=\large] at (3.65,2.80) {$\Sigma$};
  \node[text=cylBlue,anchor=west] at (5.05,4.28) {$S_R^{d-1}$};
  \node[text=cylRed,anchor=west] at (5.05,1.55) {great $S_R^1$};
  \node at (3.65,.28)
    {$\mathcal R_\Sigma=0,\qquad\Pi^\mu_{ab}=0$};
\end{scope}
\end{tikzpicture}%
}
\caption{Two cylindrical embeddings of a conformal surface defect, with ambient spatial slices shown schematically in blue.
(a) A circular cylinder in flat $\mathbb R^d$ has
$\mathring\Pi^{\,2}=1/(2R^2)$; $n$ is an outward unit normal.
(b) Ambient radial quantization maps a plane through the origin to
$\mathbb R_\tau\times S_R^1\subset\mathbb R_\tau\times S_R^{d-1}$,
with $\Pi^\mu_{ab}=0$. Both induced metrics are intrinsically flat.}
\label{fig:cylinder-embeddings}
\end{figure*}
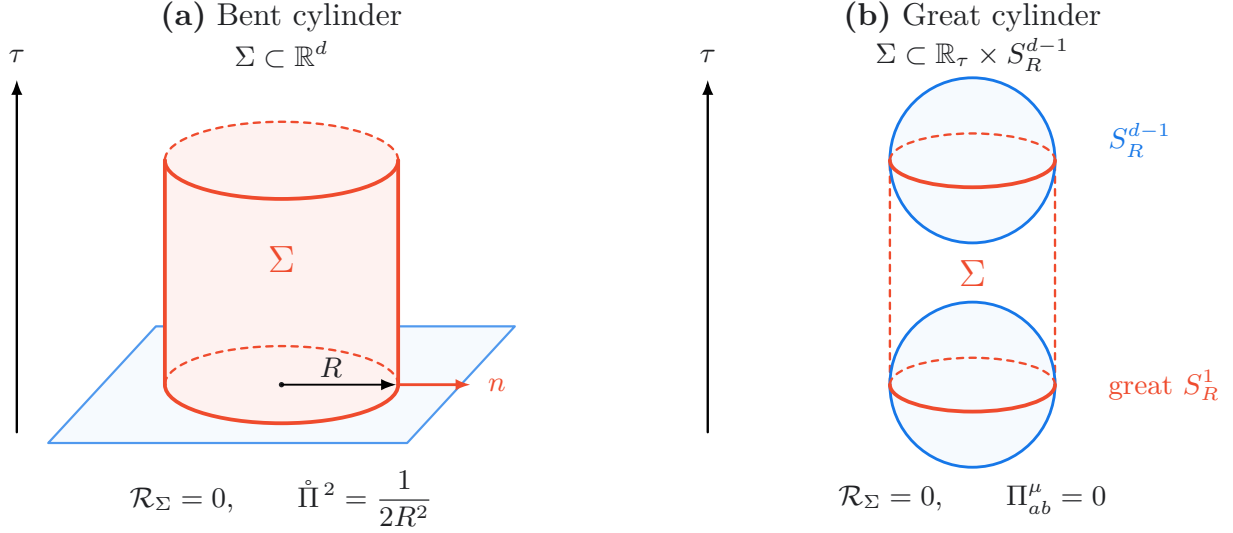

Two inequivalent cylindrical embeddings share the induced metric
$\mathrm{d}\tau^2+R^2\mathrm{d}\varphi^2$,
as illustrated in Fig.~\ref{fig:cylinder-embeddings}. A circular cylinder
$\mathbb{R}\times S_R^1\subset\mathbb{R}^d$ is extrinsically curved.
Radial quantization of the ambient theory instead maps a plane through
the origin to a totally geodesic great cylinder in
$\mathbb{R}\times S_R^{d-1}$.
For these two geometries, we find
\begin{align}
 E_{\rm bent}^{\log}
 &=-\frac{d_1}{24R}\log\frac{R}{\epsilon},
 \label{eq:mainbent}\\
 E_{\rm great}
 &=-\frac{b}{12R}.
 \label{eq:maingreat}
\end{align}
Here $b$ and $d_1$ are the defect anomaly coefficients defined below.
Equation~\eqref{eq:mainbent} fixes the scheme-independent coefficient
of the logarithmic term; the accompanying finite $1/R$ term is not
universal. Equation~\eqref{eq:maingreat} gives the finite energy in a
specified prescription: the complete flat-space dilatation charge
annihilates the defect vacuum, and the Weyl transformation is
implemented with its Wess--Zumino (WZ) functional in a fixed convention
for trivial anomalies.

Although finite local counterterms can affect ordinary Casimir
energies~\cite{HerzogHuang,Assel}, the admissible freedom here is
restricted by the defect Ward identities and
the chosen convention for trivial anomalies~\cite{JensenEntropy}. Once these conditions
are imposed, the remaining finite Weyl-invariant counterterms
give no extensive contribution on the great cylinder
and therefore cannot shift its vacuum energy.
Equation~\eqref{eq:maingreat} is thus uniquely determined within this
class of renormalization schemes.
This ordinary defect Casimir energy is distinct from its
supersymmetric counterpart, which involves additional background
fields and different anomaly combinations~\cite{Bobev,Huang}.

\textit{Anomaly and observable.---}Consider a parity-even conformal
surface defect $\Sigma$ in a $d$-dimensional CFT with $d\geq4$ and
no additional background sources. Its induced metric is $\gamma_{ab}$,
and its second fundamental form is $\Pi^\mu_{ab}$, with trace
$\Pi^\mu=\gamma^{ab}\Pi^\mu_{ab}$ and traceless part
$\mathring\Pi^\mu_{ab}
=\Pi^\mu_{ab}-\frac12\gamma_{ab}\Pi^\mu$.
For the Euclidean functional defined below, write
$\delta_\omega W_{\cD}=\int\dd^dx\sqrt g\,\omega\mathcal A_{\cD}$.
We normalize the Weyl-response density as~\cite{GrahamWitten,JensenEntropy}
\begin{equation}
 \mathcal A_{\cD}
 =-\frac{\delta_\Sigma}{24\pi}
 \bigl[b\Ric_\Sigma+d_1\Ksq-d_2\Wpull\bigr],
 \label{eq:anomaly}
\end{equation}
where $\Ric_\Sigma$ is the intrinsic scalar curvature,
$\Ksq=\mathring\Pi^\mu_{ab}\mathring\Pi_\mu^{ab}$,
$\Wpull$ is the tangential contraction of the ambient Weyl tensor,
and $\delta_\Sigma$ is the covariant delta function supported on
the defect. Following Ref.~\cite{JensenEntropy}, we choose finite local
counterterms to remove terms involving normal derivatives of
$\delta_\Sigma$ from the Weyl response, bringing it to the
distributional form in Eq.~\eqref{eq:anomaly}.
This choice fixes our trivial-anomaly convention while retaining
the freedom to add finite Weyl-invariant counterterms.

We define the defect generating functional and vacuum energy by
\begin{equation}
 W_{\cD}=-\log\frac{Z[g;\cD]}{Z[g]},\qquad
 E_{\cD}=\lim_{\beta\to\infty}\frac{W_{\cD}}{\beta},
 \label{eq:definition}
\end{equation}
where $\beta$ is the Euclidean time interval.
Our Euclidean flux tensor obeys
$\delta W_{\cD}=-\frac12\int\dd^dx\sqrt g\,
\langle T^{\mu\nu}\rangle_{\rm def}\delta g_{\mu\nu}$ at fixed embedding,
so $\langle T^\mu{}_{\mu}\rangle_{\rm def}=-\mathcal A_{\cD}$.
In particular, a round defect sphere of radius $a$ in flat space has
$W_{\cD}(S_a^2)|_{\log}=-(b/3)\log(a/\epsilon)$.

We subtract the defect tension and prepare the conformal vacuum
at the temporal ends.
The partition-function ratio cancels defect-independent bulk
contributions, including bulk anomaly terms, while retaining
all defect-induced ambient polarization.
The Euler term is type A, whereas the two Weyl-invariant
densities are type B~\cite{DeserSchwimmer,Asnin}.

\textit{Bending isolates the displacement anomaly.---}For the
circular embedding
$X^\mu=(\tau,R\cos\varphi,R\sin\varphi,0,\ldots)$,
both the intrinsic and ambient curvatures vanish, while
\begin{equation}
 \Ksq=\frac{1}{2R^2},\qquad
 \int_\Sigma\sqrt\gamma\,\Ksq=\frac{\pi\beta}{R}.
 \label{eq:bentgeometry}
\end{equation}
Only the $d_1$ term contributes to the integrated anomaly,
yielding Eq.~\eqref{eq:mainbent}.

The same argument applies to any static defect
$\R\times C\subset\R^d$, where $C$ is a smooth closed spatial
curve. 
In arc-length coordinates, the only nonzero component
of the second fundamental form is
$\Pi^\mu_{ss}=\kappa_C^\mu$, the curvature vector of $C$;
hence $\Ric_\Sigma=0$ and $\Ksq=\kappa_C^2/2$.
The ambient Weyl tensor vanishes for every such curve, so the
$d_2$ term is also absent.
Integrating the anomaly and dividing by $\beta$ therefore gives
\begin{equation}
 E_C^{\log}
 =-\frac{d_1}{48\pi}
 \left(\oint_C\dd s\,\kappa_C^2\right)
 \log\frac{\ell_C}{\epsilon},
 \label{eq:curveenergy}
\end{equation}
where $\ell_C$ is the length scale of the curve.
The coefficient scales as $\ell_C^{-1}$; shape-dependent finite
terms are not fixed by the anomaly.

The coefficient $d_1$ is fixed by the displacement two-point
function on the planar defect~\cite{Billo,BianchiRenyi,DrukkerNonlinear}:
\begin{equation}
 \langle D_i(y)D_j(0)\rangle
 =\frac{C_D\delta_{ij}}{|y|^6},\qquad
 d_1=\frac{3\pi^2}{4}C_D.
 \label{eq:displacement}
\end{equation}

Thus the planar two-point coefficient $C_D$ determines the
logarithmic bending energy for any smooth $C$:
the coefficient of $\log(\ell_C/\epsilon)$ is
$-(\pi C_D/64)\oint_C\dd s\,\kappa_C^2$.
Reflection positivity implies $C_D\geq0$, making this coefficient
nonpositive.

The finite ambiguity can be exhibited explicitly for the circular
cylinder: the Weyl-invariant counterterm
$\lambda_1\int_\Sigma\sqrt\gamma\,\Ksq$ shifts its energy by
$\pi\lambda_1/R$. The same bending analysis,
Eqs.~\eqref{eq:bentgeometry}--\eqref{eq:displacement},
also applies to two-dimensional defects in $d=3$, where the
ambient Weyl tensor vanishes identically and the $d_2$ term
is absent.

For a free field check, consider a canonically normalized
conformal scalar in six dimensions and the defect
$\cD_\lambda=\exp(\lambda\int_\Sigma\sqrt\gamma\,\phi)$.
The propagator in six-dimensional flat space is $G_6(x)=1/(4\pi^3|x|^4)$, so Gaussian
integration gives
$W_{\cD}=-(\lambda^2/2)\int_{\Sigma\times\Sigma}
\dd A\,\dd A'\,G_6(X-X')$,
where $\dd A$ is the induced area element.
Evaluating this integral on the bent cylinder yields
\begin{equation}
 E_{\phi,\rm bent}^{\log}
 =-\frac{\lambda^2}{32\pi R}\log\frac{R}{\epsilon},
 \qquad d_{1,\phi}=\frac{3\lambda^2}{4\pi},
 \label{eq:scalarbent}
\end{equation}
in agreement with Eq.~\eqref{eq:mainbent}.
Independently, the planar displacement operator
$D_i=\lambda\partial_i\phi$ gives $C_D=\lambda^2/\pi^3$,
which reproduces the same $d_{1,\phi}$ through
Eq.~\eqref{eq:displacement}.
A complementary holographic calculation with a minimal probe
brane also reproduces Eq.~\eqref{eq:mainbent}, with
$d_1=6\pi T_{\rm br}L^3$.
Details of these checks are given in Appendices~\ref{app:bent} and~\ref{app:probe}, respectively.

\textit{The complete dilatation charge.---}Ambient radial quantization maps
a defect plane through the origin to a great cylinder:
\begin{equation}
 \dd s_{\rm flat}^2=e^{2\tau/R}
 \bigl(\dd\tau^2+R^2\dd\Omega_{d-1}^2\bigr),
 \qquad r=Re^{\tau/R}.
 \label{eq:radialmap}
\end{equation}
Write $q=d-2>1$ and
$\dd\Omega_{d-1}^2=\dd\theta^2+\cos^2\theta\,\dd\varphi^2
+\sin^2\theta\,\dd\Omega_{q-1}^2$.
The defect lies at $\theta=0$. We first examine the bulk energy momentum tensor outside a tube cutoff $u:=\sin\theta>x$, before including
the anomalous Weyl contribution. In flat space, the Cartesian indices $a,b=1,2$ and
$i,j=1,\ldots,q$ label directions parallel and transverse
to the defect, respectively.
The planar one-point function is~\cite{Billo,JensenEntropy}
\begin{align}
 \langle T_{ab}\rangle&=-\frac{H}{\rho^d}\delta_{ab},\qquad
 \langle T_{ai}\rangle=0,\notag\\
 \langle T_{ij}\rangle&=\frac{H}{(q-1)\rho^d}
 \bigl(3\delta_{ij}-d n_i n_j\bigr).
 \label{eq:onepoint}
\end{align}
Here $\rho$ is transverse distance, $n_i=x_i/\rho$, and
\begin{equation}
 \mathcal N:=2\pi\Omega_{q-1}H
 =\frac{d-3}{3(d-1)}d_2,
 \label{eq:d2normalization}
\end{equation}
with $\Omega_m=\operatorname{vol}(S^m)$.
The homogeneous Weyl transform of Eq.~\eqref{eq:onepoint}
followed by analytic continuation gives
\begin{equation}
 E_{\rm out}(x)=\frac{\mathcal N}{R}\int_x^1\frac{\dd u}{u^3}
 =\frac{\mathcal N}{2R}(x^{-2}-1).
 \label{eq:charge}
\end{equation}
An area counterterm on the tube cancels the
leading divergence. Keeping its actual area gives
\begin{equation}
 E_{\rm area}(x)=-\frac{\mathcal N}{2R}\frac{\sqrt{1-x^2}}{x^2},
 \quad
 \lim_{x\to0}(E_{\rm out}+E_{\rm area})
 =-\frac{\mathcal N}{4R}.
 \label{eq:tubearea}
\end{equation}

The residual $d_2$ term is incompatible with the homogeneous
Weyl transform of the complete dilatation charge.
Indeed, the preserved dilatation generator annihilates the
planar defect vacuum, $Q_{\rm dil}|0_{\cD}\rangle=0$.
Its homogeneous Weyl transform therefore contributes
$E_{\rm hom}=-\langle Q_{\rm dil}\rangle/R=0$
to the cylinder energy.
The exterior integral, however, includes only the
separated-point stress tensor.
The Ward identity instead constrains the complete
distributional tensor, including defect-supported terms;
subtracting the exterior divergence alone does not supply
these missing contributions.

The missing ingredient is the local response of the regulator.
In the flat frame, imposing a hard transverse cutoff gives
\begin{equation}
 \partial_\mu\!\left[
 \Theta(\rho-\epsilon_{\rm f})\langle T^{\mu i}\rangle\right]
 =-H\epsilon_{\rm f}^{-d}n^i\,
 \delta(\rho-\epsilon_{\rm f}).
 \label{eq:brokenward}
\end{equation}
The cutoff thus introduces an artificial force localized on
the tube wall and also violates conservation of the truncated
dilatation current.~\footnote{Similar issues were discussed in~\cite{Lanzetta:2025xfw}.}
Subtracting an integrated power divergence does not, by itself,
restore these local Ward identities.
We therefore supplement the regulated one-point function
with a local shell contribution. Fixing the regulated radial
pressure, conservation and tracelessness determine the remaining
shell stresses in the flat vacuum.
The construction, together with its smooth-cutoff version,
is given in Appendix~\ref{app:shell}.
On a radial sphere of radius $r$, with
$x=\epsilon_{\rm f}/r$, direct integration yields
\begin{equation}
 Q_{\rm shell}(x)
 =\frac{\mathcal N}{2}\frac{1-x^2}{x^2}
 =-Q_{\rm out}(x).
 \label{eq:shellcharge}
\end{equation}
Transporting the tube and its shell completion together under
the Weyl map therefore gives a vanishing homogeneous energy,
$E_{\rm out}+E_{\rm shell}=0$.
The shell contribution already includes the subtraction of
the power divergence.
If the area counterterm in Eq.~\eqref{eq:tubearea} has already
been included, the remaining correction is
$E_{\rm shell}-E_{\rm area}$, which approaches
$+\mathcal N/(4R)$ as the cutoff is removed.
This correction cancels the apparent finite $d_2$ contribution.
The cancellation follows from this local shell construction
and reproduces the independently required vacuum Ward identity.

\textit{The Wess--Zumino contribution.---}We derive the anomalous vacuum energy from the Wess--Zumino (WZ) functional~\cite{HerzogHuang,HerzogHuangJensen}.
For a Weyl rescaling $g_{\mu\nu}\to e^{2\omega}g_{\mu\nu}$,
define $\mathcal S_b[\omega;\gamma]$ as the Euler contribution
to $W_{\cD}[e^{2\omega}g]-W_{\cD}[g]$.
Integrating the anomaly gives
\begin{align}
 \mathcal S_b[\omega;\gamma]
 =-\frac{b}{24\pi}\biggl[
 &\int_\Sigma\dd^2y\,\sqrt\gamma\,
 \bigl(\omega\Ric_\Sigma+(\nabla\omega)^2\bigr)
 \notag\\
 &+2\int_{\partial\Sigma}\dd s\,k\,\omega
 \biggr],
 \label{eq:eulerwz}
\end{align}
where $k$ is the geodesic curvature of the boundary, defined
with the outward normal.
The boundary term supplies the usual Euler completion.

To extract the energy, introduce a constant lapse $N$:
\begin{equation}
 \dd s_N^2=N^2\dd\tau^2+R^2\dd\Omega_{d-1}^2,
 \qquad
 \dd s_{\Sigma,N}^2=N^2\dd\tau^2+R^2\dd\varphi^2.
 \label{eq:lapsefamily}
\end{equation}
With $\omega=\tau/R$, the Weyl-rescaled metric
$\widetilde g_N=e^{2\omega}g_N$ is flat at $N=1$.
We vary $N$ at fixed $\omega$, $R$, and coordinate duration
$\beta$.
On the cylindrical segment, $\Ric_\Sigma=0$, and the
constant-$\tau$ end circles have $k=0$.
Since $(\nabla\omega)^2=1/(N^2R^2)$, the extensive term is
\begin{equation}
 \mathcal S_b[\omega;\gamma_N]\big|_\beta
 =-\frac{b\beta}{12RN}.
 \label{eq:wzlapse}
\end{equation}
The vacuum Ward identity makes the extensive lapse response
of $W_{\cD}[\widetilde g_N]$ vanish at $N=1$
(Appendix~\ref{app:euler}).
Thus, in the Euler sector,
$W_{\cD}[g_N]=W_{\cD}[\widetilde g_N]-\mathcal S_b$
gives
\begin{equation}
 E_b
 =-\lim_{\beta\to\infty}\frac{1}{\beta}
   \left.\frac{\partial\mathcal S_b}{\partial N}\right|_{N=1}
 =-\frac{b}{12R}.
 \label{eq:wz}
\end{equation}
The temporal cuts are sewn to the vacuum preparations
with the flat-space Ward normalization fixed above;
no independent physical boundary conditions are imposed.

Both type-B densities vanish throughout the Weyl orbit of
the constant-lapse family.
Their WZ functionals therefore have vanishing lapse derivatives
and contribute no integrated vacuum energy.
Combining the completed homogeneous charge with
Eq.~\eqref{eq:wz} gives
\begin{equation}
 E_{\rm great}
 =-\frac{\langle Q_{\rm dil}\rangle}{R}+E_b
 =-\frac{b}{12R}.
 \label{eq:greatresult}
\end{equation}

This result is unique within the stated Ward normalization
and trivial-anomaly convention.
The remaining finite Weyl-invariant counterterms cannot shift
the energy: $\int_\Sigma\sqrt\gamma\,\Ksq$ and
$\int_\Sigma\sqrt\gamma\,\Wpull$ vanish throughout the
constant-lapse family, while the intrinsic Euler term does
not affect the extensive contribution.
A finite term
$\lambda_{\Ric}\int_\Sigma\sqrt\gamma\,\Ric[g]$
would instead shift the energy by
$2\pi\lambda_{\Ric}(d-1)(d-2)/R$.
Its Weyl variation changes the trivial-anomaly representative
and is therefore incompatible with the convention fixed in
Eq.~\eqref{eq:anomaly}.
For $d=3$, the separated-point planar ambient
stress vanishes; the same Euler argument gives Eq.~\eqref{eq:maingreat}
with $d_2$ absent.

\textit{Free-field and holographic examples.---}We next examine
the finite great-cylinder energy. The scalar defect introduced
in Eq.~\eqref{eq:scalarbent} is now supported on the totally
geodesic surface
$\Sigma=\mathbb R_\tau\times S_R^1\subset
\mathbb R_\tau\times S_R^5$.
We regulate its Gaussian double integral by transverse point
splitting between the defect and an auxiliary displaced
surface~\cite{Gustavsson}.
In the angular coordinates introduced above, the original
surface lies at $\theta=0$, while the displaced surface lies
at $\sin\theta=x$ and a fixed location on the transverse $S^3$.
The auxiliary circle therefore has circumference $2\pi Rc$ with $c=\sqrt{1-x^2}$,
and the displaced surface has area $A_x=2\pi\beta Rc$.
With $\epsilon=Rx$, the area counterterm
$W_\phi^{\rm ct}=\lambda^2A_x/(8\pi^2\epsilon^2)$
exactly cancels the term proportional to $\beta$ in the
bare Gaussian functional, yielding
$E_{\phi,\rm great}=0$
(Appendix~\ref{app:gaussian}).
An independent Gaussian calculation for a round $S^2$ defect
in flat space has no logarithmic divergence, so $b_\phi=0$.
Although $d_{2,\phi}=\lambda^2/(2\pi)$ is nonzero, no finite
energy remains in this prescription.

For a six-dimensional nonchiral two-form with action
$S_B=(12g^2)^{-1}\int\dd^6x\sqrt g\,
H_{\mu\nu\rho}H^{\mu\nu\rho}$, $H=\dd B$, consider the
Wilson surface $\exp(i e\int_\Sigma B)$ and set
$\kappa=e^2g^2$.
Using the same transverse point splitting, with the area
counterterm evaluated on the displaced surface, we find
a nonzero finite energy for the Wilson surface.
The logarithmic term for a round spherical defect in flat
space independently determines the $b$ coefficient~\cite{DrukkerSurface}:
\begin{equation}
 E_{B,\rm great}=-\frac{\kappa}{8\pi R},
 \qquad b_B=\frac{3\kappa}{2\pi}.
 \label{eq:twoformresult}
\end{equation}
Thus $E_{B,\rm great}=-b_B/(12R)$.
These Gaussian results are consistency checks in the stated
subtraction scheme. A further understanding of this consistency requires
matching this subtraction prescription to the
specified trivial-anomaly convention and vacuum Ward normalization (Appendix~\ref{app:gaussian}).

A complementary calculation in holography uses a constant-tension probe
brane with a totally geodesic $\mathrm{AdS}_3$ worldvolume.
Define $\cT=T_{\rm br}L^3$, where $L$ is the AdS radius.
The renormalized on-shell action gives the leading defect
contribution to $W_{\cD}$.
Regulating the brane at the dimensionless global radius
$v=v_c$ and adding the local boundary-area counterterm gives
\begin{equation}
 E_{\rm br}
 =\frac{\pi\cT}{R}\lim_{v_c\to\infty}
 \left[v_c^2-v_c\sqrt{1+v_c^2}\right]
 =-\frac{\pi\cT}{2R}.
 \label{eq:probeenergy}
\end{equation}
The spherical defect independently gives $b=6\pi\cT$,
reproducing Eq.~\eqref{eq:maingreat}.
This model also has $d_1=d_2=b$ at classical probe
order~\cite{GrahamWitten,JensenEntropy}.
For a fundamental M2-brane in $\mathrm{AdS}_7\times S^4$,
$\cT=2N/\pi$~\cite{KlebanovTseytlin1996,Tseytlin2000}, giving
$E_{\rm great}=-N/R$ and
$E_{\rm bent}^{\log}=-N\log(R/\epsilon)/(2R)$
for the two embeddings at this order.
Details are given in Appendix~\ref{app:probe}.

\textit{Discussions.---}The two embeddings probe complementary
defect data: the bending logarithm determines $C_D$, while the
great-cylinder energy determines $b$ in the stated prescription.
Existing anomaly results for critical $O(N)$ surface defects
as well as other surface defects therefore yield finite-size
predictions~\cite{GiombiLiu,Trepanier,Diatlyk,Monodromy,Hyperbolic}.
For four-dimensional replica defects, the bending coefficient
also probes the displacement data governing R\'enyi shape
dependence~\cite{LewkowyczPerlmutter,BianchiRenyi}. It would be interesting to extend our analysis to surfaces with a crease~\cite{DrukkerTrepanier2021BPS,DrukkerTrepanier2022Crease}, generalizing
the cusp geometry of line defects~\cite{DrukkerForini,Cuomo:2024psk}.

For a local, unitary defect RG flow with the ambient CFT held
fixed, the $b$-theorem~\cite{JensenRG,Wang} orders the great cylinder energies at
the conformal endpoints at equal $R$ with the same
renormalization prescription.
The energy itself has no universal sign, since $b<0$ occurs
even in simple examples of defects~\cite{KrishnanMetlitski,Georgiou}.

The absence of $d_2$ from the great-cylinder energy is specific
to this observable. For a ball of radius $\ell$ centered on
a planar defect, the defect entanglement entropy contains
$\frac13\bigl(b-\frac{d-3}{d-1}d_2\bigr)
\log(\ell/\epsilon)$~\cite{JensenHolography,JensenEntropy,Kobayashi:2018lil}.
Its modular generator differs from the global radial
Hamiltonian~\cite{Casini}, while supersymmetric anomaly
relations~\cite{Huang,Chalabi:2020iie} concern a different
vacuum observable.

In $d=3$, fuzzy spheres offer a prospective numerical
test~\cite{FuzzySphere,FuzzySurface,Feng:2026iii,Dedushenko:2024nwi}.
Two decoupled hemispheres with boundary conditions $A$ and
$B$ define an equatorial defect with $b=b_A+b_B$.
Wavefunction overlaps provide independent estimates of
$b_A$ and $b_B$. Comparing the combined hemisphere ground-state energy
with that of the full sphere then requires retaining
additive Hamiltonian constants and matching local
counterterms to the stated trivial-anomaly convention.
The free-scalar hemisphere calculation in Appendix~\ref{app:hemisphere-scalar}
illustrates this matching.

Two-dimensional Rydberg arrays offer a potential experimental
route to the bending relation through their Ising criticality
and local detuning control~\cite{RydbergIsing,RydbergLocal}.
A circular detuning pattern could realize a spatial line
defect with a two-dimensional worldsheet.
If the defect reaches a conformal fixed point in a critical
bulk, its energy contains $-vd_1\log(R/a)/(24R)$, where $a$
is the microscopic cutoff and $v$ is the emergent bulk
velocity. In a regime with negligible lattice and
outer-boundary corrections, a radius scan could extract
this coefficient by subtracting local power terms and
separating the logarithm from the nonuniversal finite
$1/R$ contribution.

\begin{acknowledgments}
We thank Yin-Chen He, Fedor Popov, Slava Rychkov, Amit Sever, Zimo Sun, Yifan Wang, Siwei Zhong and our group members at Fudan University for helpful discussion.
This work is supported by NSFC grant 12375063 and also supported by Shanghai Oriental Talents
Program.

\end{acknowledgments}

\onecolumngrid
\clearpage
\begin{center}\textbf{APPENDICES}\end{center}
\twocolumngrid
\appendix
\setcounter{secnumdepth}{2}
\numberwithin{equation}{section}
\renewcommand{\theequation}{\thesection\arabic{equation}}

\section{Gaussian check of the bending logarithm}
\label{app:bent}

For the six-dimensional free scalar, regulate the Gaussian
double integral by the ambient distance cutoff
$|X-X'|\geq\epsilon$.
Fix one insertion on the circular cylinder of radius $R$
and denote the Euclidean-time and angular separations by
$t$ and $\varphi$.
Setting $s=2R\sin(\varphi/2)$, with
$-\pi\leq\varphi\leq\pi$, gives
\[
 |X-X'|^2=t^2+s^2,
 \qquad
 \dd A'=\frac{\dd t\,\dd s}{\sqrt{1-s^2/(4R^2)}}.
\]
The first two terms in the local expansion
\begin{equation}
 \frac{1}{\sqrt{1-s^2/(4R^2)}}
 =1+\frac{s^2}{8R^2}+O(s^4/R^4)
 \label{eq:appbentjac}
\end{equation}
generate the power and logarithmic divergences, respectively.
Using $G_6=1/[4\pi^3(t^2+s^2)^2]$ and taking the
Euclidean time interval $\beta$ to infinity, we obtain
\begin{align}
 \lim_{\beta\to\infty}\frac1\beta
 \int_\Sigma\dd A\int_\Sigma\dd A'\,G_6
 &=\frac{R}{2\pi\epsilon^2}
   +\frac{1}{16\pi R}\log\frac{R}{\epsilon}
   \notag\\
 &\quad+\frac{c_{\rm fin}}{R}
   +O\!\left(\frac{\epsilon^2}{R^3}\right),
 \label{eq:appbent}
\end{align}
as $\epsilon/R\to0$.
Here $c_{\rm fin}$ is a regulator-dependent finite constant.
Multiplication by $-\lambda^2/2$ reproduces
Eq.~\eqref{eq:scalarbent}.
An angular-separation cutoff changes the power divergence
but leaves the logarithmic coefficient unchanged.

\section{Tube subtraction and the shell contribution}
\label{app:shell}

We compute the homogeneous contribution to the cylinder
energy, with $q=d-2>1$, $\Omega_m=\operatorname{vol}(S^m)$,
and $\mathcal N=2\pi\Omega_{q-1}H$.
The anomalous Weyl contribution is treated separately.

\subsection{Area subtraction}

The stationary tube wall lies at $u=\sin\theta=x$.
Writing $a=Rx$ and $c_x=\sqrt{1-x^2}$, its volume over
a Euclidean duration $\beta$ is
\begin{equation}
 \int_{\partial N_x}\dd^{d-1}y\,\sqrt h
 =2\pi\beta\,\Omega_{q-1}R^q c_xx^{q-1},
 \label{eq:app-tubearea}
\end{equation}
where $h$ is the induced metric.
The area counterterm
\begin{equation}
 I_{\rm area}
 =-\frac{H}{2a^{q+1}}
   \int_{\partial N_x}\dd^{d-1}y\,\sqrt h
 \label{eq:app-areaaction}
\end{equation}
gives
\begin{align}
 E_{\rm area}=\frac{I_{\rm area}}\beta
 &=-\frac{\mathcal N}{2R}\frac{c_x}{x^2}\notag\\
 &=-\frac{\mathcal N}{2R}x^{-2}
   +\frac{\mathcal N}{4R}
   +O\!\left(\frac{x^2}{R}\right).
 \label{eq:app-areatail}
\end{align}
The cutoff $a=R\sin\theta$ is a coordinate radius;
the geodesic tube radius is $R\theta$.

\subsection{Local Ward completion}

In flat space, let $y^a$ ($a=1,2$) and $z^i$
($i=1,\ldots,q$) denote parallel and transverse coordinates.
Set $X^\mu=(y^a,z^i)$, $\rho=|z|$, and $n_i=z_i/\rho$.
Choose a smooth cutoff $F(\rho)$ that vanishes for
$\rho<\epsilon_{\rm f}$ and equals one for
$\rho>2\epsilon_{\rm f}$.
The truncated one-point function $F\langle T_{\mu\nu}\rangle$
is traceless but obeys
\begin{align}
 \partial_\mu(F\langle T^{\mu i}\rangle)
 &=-H\rho^{-d}F'n^i,\notag\\
 \partial_\mu j_F^\mu&=-H\rho^{1-d}F',
 \qquad
 j_F^\mu=X_\nu F\langle T^{\mu\nu}\rangle .
 \label{eq:app-cutoffcurrent}
\end{align}
The cutoff produces a wall force and violates conservation
of the truncated dilatation current.
An integrated power subtraction does not restore these
local Ward identities.

Write the completed one-point function as
$\mathcal T_{\mu\nu}=F\langle T_{\mu\nu}\rangle
+\mathcal T_{\mu\nu}^{\rm shell}$.
Planar translations and $O(2)\times O(q)$ rotations imply
\begin{align}
 \mathcal T_{ab}&=A(\rho)\delta_{ab},
 &\mathcal T_{ai}&=0,\notag\\
 \mathcal T_{ij}
 &=B(\rho)n_i n_j
   +C(\rho)(\delta_{ij}-n_i n_j).
 \label{eq:app-ansatz}
\end{align}
The planar vacuum has $\langle D_i\rangle=0$ and vanishing
geometric anomaly densities.
In our convention, which removes normal derivatives of
$\delta_\Sigma$ from the trace, the Ward conditions are
\begin{equation}
 B'+\frac{q-1}{\rho}(B-C)=0,
 \qquad 2A+B+(q-1)C=0.
 \label{eq:app-ward}
\end{equation}
Fixing the regulated radial pressure $B=-H\rho^{-d}F$
determines
\begin{align}
 A&=-\frac{qB+\rho B'}2
   =-H\rho^{-d}F+\frac H2\rho^{1-d}F',\notag\\
 C&=B+\frac{\rho B'}{q-1}
   =\frac{H}{q-1}
     \left(3\rho^{-d}F-\rho^{1-d}F'\right).
 \label{eq:app-shellsolution}
\end{align}
The $F'$ terms give the shell, fixed within this ansatz
before charge integration.
For $F=\Theta(\rho-\epsilon_{\rm f})$, they become
\begin{align}
 \mathcal T^{\rm shell}_{ab}
 &=\frac{H\delta_{ab}}{2\epsilon_{\rm f}^{d-1}}
   \delta(\rho-\epsilon_{\rm f}),\notag\\
 \mathcal T^{\rm shell}_{ij}
 &=-\frac{H(\delta_{ij}-n_i n_j)}
          {(q-1)\epsilon_{\rm f}^{d-1}}
   \delta(\rho-\epsilon_{\rm f}),
 \label{eq:app-sharpshell}
\end{align}
with $\mathcal T^{\rm shell}_{ai}=0$.
The shell is traceless and cancels the wall force in
Eq.~\eqref{eq:app-cutoffcurrent}.

\subsection{Complete charge and Weyl transport}

On a radial sphere $S_r^{d-1}$ in flat space, set $u=\rho/r$.
The regulated vacuum dilatation flux is~\footnote{Our convention of dilatation generator is different from Duffin's CFT lecture~\cite{SimmonsDuffin:2016gjk} by a minus sign.}
\begin{align}
 \langle Q_{\rm dil}(r)\rangle_{\epsilon_{\rm f}}
 &\equiv\int_{S_r^{d-1}}\dd S_\mu\,
       X_\nu\mathcal T^{\mu\nu}\notag\\
 &=2\pi\Omega_{q-1}r^d\int_0^1\dd u\,u^{q-1}\notag\\
 &\qquad\times
 \left[A(ru)(1-u^2)+B(ru)u^2\right].
 \label{eq:app-fullintegral}
\end{align}
Substituting Eq.~\eqref{eq:app-shellsolution} turns the
integrand into a total derivative:
\begin{align}
 &u^{q-1}\left[A(ru)(1-u^2)+B(ru)u^2\right]\notag\\
 &\qquad=-\frac12\frac{\dd}{\dd u}
       \left[u^q(1-u^2)B(ru)\right].
 \label{eq:app-totalderivative}
\end{align}
The boundary term vanishes at $u=1$ because of the factor
$1-u^2$, and at $u=0$ because $F$ vanishes near the defect.
Thus
\begin{equation}
\langle Q_{\rm dil}(r)\rangle_{\epsilon_{\rm f}}
 =-\pi\Omega_{q-1}r^d
   \left[u^q(1-u^2)B(ru)\right]_0^1=0.
 \label{eq:app-smoothzero}
\end{equation}
The regulated flux vanishes at every nonzero cutoff,
reproducing $\langle Q_{\rm dil}\rangle=0$ in the
vacuum Ward normalization.
For the sharp cutoff, set $x=\epsilon_{\rm f}/r<1$.
Using $\delta(ru-\epsilon_{\rm f})=r^{-1}\delta(u-x)$,
the two contributions are
\begin{equation}
 Q_{\rm out}=-\frac{\mathcal N}{2}(x^{-2}-1),
 \qquad
 Q_{\rm shell}=\frac{\mathcal N}{2}\frac{1-x^2}{x^2}.
 \label{eq:app-cancellation}
\end{equation}
The shell cancels both the divergent and finite parts of
the exterior flux.

Under the radial Weyl map $r=Re^{\tau/R}$, we transport
the stress distribution together with its cutoff and shell.
The relation $\partial_\tau=R^{-1}X^\mu\partial_\mu$
supplies a factor $1/R$, while the homogeneous Weyl weights
of the stress tensor and surface measure cancel.
In our Euclidean outward-flux convention, the homogeneous
cylinder energy is therefore
\begin{equation}
 E_{\rm hom}(\tau)=-\frac{\langle Q_{\rm dil}(r)\rangle_{\epsilon_{\rm f}}}R=0.
 \label{eq:app-homogeneouszero}
\end{equation}
A fixed flat-space cutoff $\epsilon_{\rm f}$ maps to a
moving cylinder tube, $a(\tau)=\epsilon_{\rm f}e^{-\tau/R}$.
To compare with the stationary tube prescription, identify
$x=\epsilon_{\rm f}/r=a/R$ on a fixed time slice.
The difference between the shell contribution and the
area subtraction is then
\begin{equation}
 E_{\rm shell}-E_{\rm area}
 =\frac{\mathcal N}{2R}\frac{c_x-c_x^2}{x^2}
 \longrightarrow\frac{\mathcal N}{4R}.
 \label{eq:app-matcharea}
\end{equation}
Adding this correction to the area-subtracted exterior
energy gives zero.

\section{WZ functional and the lapse response}
\label{app:euler}

We relate the flat-frame lapse response to the complete
dilatation charge and obtain the cylinder energy from
the Euler WZ term.
For the constant-lapse family~\eqref{eq:lapsefamily},
hold $\omega=\tau/R$, $R$, the embedding, and $\beta$ fixed.
With $r=Re^{\tau/R}$,
\begin{equation}
 \dd\widetilde s_N^2
 =e^{2\omega}\dd s_N^2
 =N^2\dd r^2+r^2\dd\Omega_{d-1}^2,
 \label{eq:app-conefamily}
\end{equation}
which is flat at $N=1$.

Our Euclidean convention is
\begin{equation}
 \delta W_{\cD}
 =-\frac12\int\dd^dX\sqrt g\,
   \langle T^{\mu\nu}\rangle_{\rm def}\delta g_{\mu\nu}.
 \label{eq:app-metricresponse}
\end{equation}
At $N=1$, $\partial_N\widetilde g_{rr}=2$.
Keeping the vacuum preparations fixed, the extensive
response is
\begin{align}
 \left.\partial_N W_{\cD}[\widetilde g_N]\right|_{N=1}
 \bigg|_\beta
 &=-\int_{r_-}^{r_+}\dd r
   \int_{S_r^{d-1}}\dd S\,
   \langle T^{rr}\rangle_{\rm def}\notag\\
 &=-\int_{r_-}^{r_+}\frac{\dd r}{r}\,
   \langle Q_{\rm dil}(r)\rangle\notag\\
 &=-\frac{\beta}{R}\langle Q_{\rm dil}\rangle=0.
 \label{eq:app-flatlapseresponse}
\end{align}
Here $\langle Q_{\rm dil}(r)\rangle
=r\int_{S_r^{d-1}}\dd S\,\langle T^{rr}\rangle_{\rm def}$
is the complete outward dilatation charge, including
the contact terms, and $\log(r_+/r_-)=\beta/R$.
The last equality uses the vacuum Ward normalization
$Q_{\rm dil}|0_{\cD}\rangle=0$, implemented by the shell
completion in Appendix~\ref{app:shell}.

For every constant $N$, the ambient metric is conformally
flat and the great defect is totally geodesic.
Both type-B densities vanish along the Weyl transformation,
as do their lapse responses.
On the defect, $\Ric[\gamma_N]=k=0$,
$\sqrt{\gamma_N}=NR$, and
$(\nabla\omega)^2=1/(N^2R^2)$.
The Euler WZ functional~\eqref{eq:eulerwz}, with its boundary
completion~\cite{HerzogHuangJensen}, gives
\begin{equation}
 \mathcal S_b[\omega;\gamma_N]\big|_\beta
 =-\frac{b\beta}{12RN}.
 \label{eq:app-wzlapse}
\end{equation}
Using $W_{\cD}[g_N]=W_{\cD}[\widetilde g_N]-\mathcal S_b$,
we therefore find
\begin{equation}
 E_{\rm great}
 =-\lim_{\beta\to\infty}\frac1\beta
   \left.\partial_N\mathcal S_b\right|_{N=1}
 =-\frac{b}{12R}.
 \label{eq:app-eulerenergy}
\end{equation}

Finite, parity-even curvature counterterms with dimensionless
coefficients must preserve Eq.~\eqref{eq:anomaly}.
The Weyl-invariant terms built from $\Ksq$ and $\Wpull$
vanish on this family, and the completed intrinsic Euler
term has no extensive contribution.
Other curvature terms can shift the energy but alter the
trivial-anomaly convention. For example,
$I_{\Ric}=\int_\Sigma\dd^2y\sqrt\gamma\,\Ric[g]$ obeys
\begin{equation}
 \delta_\sigma I_{\Ric}
 =-2(d-1)\int_\Sigma\dd^2y\sqrt\gamma\,
   \nabla_g^2\sigma,
 \label{eq:app-trivialvariation}
\end{equation}
which introduces normal-derivative contact terms in the
trace. Such shifts are excluded by the prescribed Weyl
response, leaving no further energy ambiguity within
this counterterm class.

\section{Gaussian point splitting and its finite subtraction}
\label{app:gaussian}

We regulate the Gaussian self-action on the great cylinder
in $\mathbb R_\tau\times S_R^5$ by transverse point splitting.
Using the angular coordinates introduced above, choose
\begin{equation}
 \begin{aligned}
 \Sigma_0:&\quad \theta=0,\\
 \Sigma_x:&\quad \sin\theta=x,\qquad
 \boldsymbol{\vartheta}=\boldsymbol{\vartheta}_*,
 \end{aligned}
 \label{eq:appsplit}
\end{equation}
where $0<x<1$ and $\boldsymbol{\vartheta}_*$ specifies a
fixed point on the transverse $S^3$.
Both surfaces are parametrized by $(\tau,\varphi)$.
Writing $c=\sqrt{1-x^2}$ and $\epsilon=Rx$, the auxiliary
surface has area $A_x=2\pi\beta Rc$, with $\beta$ the
Euclidean duration.
This prescription retains the Gaussian factor $1/2$ and
introduces no second physical defect.
We define
$E^{\rm bare}(x)=
\lim_{\beta\to\infty}W^{\rm bare}(x;\beta)/\beta$
before taking $x\to0$.

For the scalar insertion $\exp(\lambda\int_\Sigma\dd A\,\phi)$,
use the canonical action
$S_\phi=\frac12\int\dd^6x\sqrt g\,
[(\nabla\phi)^2+\Ric[g]\phi^2/5]$.
The Green function between the two surfaces is
\begin{equation}
 G_{\rm cyl}
 =\frac{1}{16\pi^3R^4(\cosh t-c\cos\varphi)^2},
 \qquad t=\frac{\tau-\tau'}R,
 \label{eq:appG}
\end{equation}
where $\varphi$ denotes the angular separation.
Angular integration gives
$2\pi\cosh t/(\sinh^2t+x^2)^{3/2}$; substituting
$s=\sinh t$ then yields
\begin{equation}
 \int_{-\infty}^{\infty}\dd t\int_0^{2\pi}\dd\varphi\,
 (\cosh t-c\cos\varphi)^{-2}=\frac{4\pi}{x^2}.
 \label{eq:appintegral}
\end{equation}
The Gaussian integral and prescribed area counterterm give
\begin{equation}
 E_\phi^{\rm bare}(x)=-\frac{\lambda^2c}{4\pi R x^2},
 \qquad
 W_\phi^{\rm ct}=\frac{\lambda^2 A_x}{8\pi^2\epsilon^2}.
 \label{eq:appscalar}
\end{equation}
Their extensive parts cancel at every $x$, including the
finite terms, so $E_{\phi,\rm great}^{\rm split}=0$.

Independently, a round sphere of radius $a$ in flat space,
regulated by the ambient chord cutoff $|X-X'|\geq\epsilon$,
gives
\begin{equation}
 W_\phi(S_a^2)
 =-\frac{\lambda^2a^2}{2\pi\epsilon^2}
   +\frac{\lambda^2}{8\pi}.
 \label{eq:appscalarsphere}
\end{equation}
Since the type-B densities vanish on this sphere, the
absence of a logarithm implies $b_\phi=0$.
The planar profile
$\langle\phi\rangle=\lambda/(4\pi^2\rho^2)$ gives
$H=\lambda^2/(40\pi^4)$ from the improved stress tensor
away from the defect, and hence
$d_{2,\phi}=\lambda^2/(2\pi)\ne0$.

For the nonchiral two-form, set $\kappa=e^2g^2$ and use
the flat Feynman-gauge propagator
\begin{align}
 &\langle B_{\mu\nu}(X)B_{\rho\sigma}(X')\rangle\notag\\
 &\quad=g^2(\delta_{\mu\rho}\delta_{\nu\sigma}
          -\delta_{\mu\sigma}\delta_{\nu\rho})G_6(X-X'),
\end{align}
where $G_6(Y)=1/(4\pi^3|Y|^4)$.
With $\int_\Sigma B=\frac12\int_\Sigma
\dd X^\mu\wedge\dd X^\nu B_{\mu\nu}$, the insertion
$\exp(i e\int_\Sigma B)$ gives the opposite Gaussian sign.
We pull the propagator back from the Weyl-related flat
radial frame, $g_{\rm flat}=e^{2\tau/R}g_{\rm cyl}$.
Writing $B^{(0)}$ and $B^{(x)}$ for the surface pullbacks,
the oriented tangent contraction gives
\begin{equation}
 \bigl\langle
 B^{(0)}_{\tau\varphi}B^{(x)}_{\tau'\varphi'}
 \bigr\rangle
 =g^2R^2c^2G_{\rm cyl}.
 \label{eq:appBpullback}
\end{equation}
Equation~\eqref{eq:appintegral} therefore yields
\begin{equation}
 E_B^{\rm bare}(x)=\frac{\kappa c^2}{4\pi R x^2},
 \qquad
 W_B^{\rm ct}=-\frac{\kappa A_x}{8\pi^2\epsilon^2}.
 \label{eq:appB}
\end{equation}
Since $(c^2-c)/x^2=-c/(1+c)\to-1/2$, this prescription
gives $E_{B,\rm great}^{\rm split}=-\kappa/(8\pi R)$.
With $b_B=3\kappa/(2\pi)$, this is consistent with
$E_{B,\rm great}^{\rm split}=-b_B/(12R)$.

A complete test further
requires matching this subtraction prescription to the
specified trivial-anomaly convention.

\section{Probe-brane calculations}
\label{app:probe}

Consider a three-dimensional pure-tension brane in Euclidean
$\mathrm{AdS}_{d+1}$ of radius $L$, and define
$\cT=T_{\rm br}L^3$. For cylindrical defects, we keep the
terms proportional to the boundary Euclidean duration $\beta$.

For a circular cylinder of radius $R$ in a flat boundary
background, the rotationally invariant embedding $r=r(z)$
in Poincar\'e coordinates has action
\begin{equation}
 W_{\rm br}^{\rm reg}
 =2\pi\beta\cT\int_\epsilon^{z_*}
 \frac{\dd z}{z^3}\,r(z)\sqrt{1+r'(z)^2},
 \label{eq:appbrane}
\end{equation}
where $r(0)=R$, $r'=\dd r/\dd z$, and $z_*$ is the
interior endpoint. The extremality equation,
$r''/(1+r'^2)-3r'/z-1/r=0$, gives
\begin{align}
 r(z)&=R-\frac{z^2}{4R}
 +O\!\left(\frac{z^4}{R^3}\log\frac Rz\right),\notag\\
 r\sqrt{1+r'^2}&=R-\frac{z^2}{8R}
 +O\!\left(\frac{z^4}{R^3}\log\frac Rz\right).
 \label{eq:appbrane-expansion}
\end{align}
The logarithmic energy is therefore
$E_{\rm br,bent}^{\log}
=-\pi\cT\log(R/\epsilon)/(4R)$,
consistent with $d_1=6\pi\cT$~\cite{GrahamWitten,JensenEntropy}.
Its coefficient depends only on the near-boundary expansion.

For the great cylinder in $\mathbb R_\tau\times S_R^{d-1}$,
the totally geodesic brane has induced metric
\begin{equation}
 \dd s_{\rm wv}^2=L^2\left[
 (1+v^2)\frac{\dd\tau^2}{R^2}
 +\frac{\dd v^2}{1+v^2}+v^2\dd\varphi^2\right],
 \label{eq:appglobal}
\end{equation}
where $v\geq0$, $\varphi\sim\varphi+2\pi$, and $v=0$
is the smooth center. A radial cutoff at $v=v_c$ gives
$W_{\rm br}^{\rm reg}/\beta=\pi\cT v_c^2/R$.
The local area counterterm is
\begin{equation}
 I_{\rm ct}
 =-\frac{T_{\rm br}L}{2}\int\dd^2y\sqrt h
 =-\frac{\pi\beta\cT}{R}v_c\sqrt{1+v_c^2},
 \label{eq:app-probect}
\end{equation}
with $h$ the induced metric at the cutoff. Thus
\begin{align}
 \frac{W_{\rm br}^{\rm reg}+I_{\rm ct}}{\beta}
 &=\frac{\pi\cT}{R}
   \left[v_c^2-v_c\sqrt{1+v_c^2}\right]\notag\\
 &=-\frac{\pi\cT}{2R}
   +O\!\left(\frac{\cT}{Rv_c^2}\right).
 \label{eq:app-fgenergy}
\end{align}
The finite term comes from retaining the full cutoff area.

Independently, the $\mathrm H^3$ filling of a spherical
defect $S_a^2$ has regulated volume
\begin{equation}
 V_{\rm reg}=L^3[\pi\sinh(2\eta_c)-2\pi\eta_c],
 \qquad \cosh\eta_c=\frac a\epsilon,
 \label{eq:app-probesphere}
\end{equation}
with Poincar\'e cutoff $z=\epsilon$.
Since the area counterterm contains no logarithm,
$W_{\rm br}(S_a^2)|_{\log}=-2\pi\cT\log(a/\epsilon)$.
Matching to $-(b/3)\log(a/\epsilon)$ gives
$b=6\pi\cT$, verifying $E_{\rm br,great}=-b/(12R)$.

\section{Free-scalar hemisphere check}
\label{app:hemisphere-scalar}

Consider a real conformal scalar on
$M_\beta=S^1_\beta\times HS_R^2$, with periodic Euclidean
time and $P=-\nabla^2+\Ric[g]/8$.
Set $\chi=-1$ for Dirichlet and $\chi=+1$ for conformal
Robin boundary conditions, $(\nabla_n+K/4)\phi=0$,
where $n$ is the outward unit normal and $K$ is the trace
of the extrinsic curvature.
At the equatorial boundary $\Sigma=\partial M_\beta$,
$K_{ab}=0$, so Robin reduces to Neumann.
The frequencies and multiplicities are
\begin{equation}
 \omega_l=\frac{l+\tfrac12}{R},\qquad
 g_l^\chi=l+\frac{1+\chi}{2},\qquad l=0,1,\ldots .
 \label{eq:hs-scalar-spectrum}
\end{equation}
For $W_\chi^\zeta=\tfrac12\log\det_\zeta P$, the vacuum
energy $E_\chi^\zeta=\lim_{\beta\to\infty}W_\chi^\zeta/\beta$
is therefore
\begin{equation}
 E_\chi^\zeta=\frac1{2R}\left[
 \zeta_{\rm H}\!\left(-2,\tfrac12\right)
 +\frac\chi2\zeta_{\rm H}\!\left(-1,\tfrac12\right)\right]
 =\frac\chi{96R},
 \label{eq:hs-scalar-zeta}
\end{equation}
where $\zeta_{\rm H}$ is the Hurwitz zeta function.
The full sphere has $E_{S^2}^\zeta=0$, while the independent
boundary-anomaly result is $b_\chi=\chi/16$~\cite{JensenEntropy}.

The spectral prescription must be matched to the stated
trivial-anomaly convention.
Let $n^\mu$ be the outward unit normal and
$K=\gamma^{ab}K_{ab}$, with
$K_{ab}=\gamma_a{}^\mu\gamma_b{}^\nu\nabla_\mu n_\nu$.
Define
$\sigma_n=n^\mu\nabla_\mu\sigma$ and
$\sigma_{nn}=n^\mu n^\nu\nabla_\mu\nabla_\nu\sigma$.
For $P=-\nabla^2+\Ric[g]/8$, the smeared heat kernel gives
$\delta_\sigma W_\chi^\zeta=-a_3(\sigma,P)$.

On a general smooth boundary, the normal-derivative terms
are~\cite{VassilevichHeatKernel}
\begin{equation}
 \delta_\sigma W_\chi^\zeta\supset
 -\frac1{256\pi}\int_\Sigma\dd^2y\sqrt\gamma\,
 \left[4\chi\sigma_{nn}+(4\chi-1)K\sigma_n\right].
 \label{eq:hs-scalar-trivial}
\end{equation}
For $K_{ab}=0$, only the first term remains.
To cancel both terms, add
\begin{equation}
 W_\chi^{\rm ct}
 =-\frac1{256\pi}\int_\Sigma\dd^2y\sqrt\gamma\,
 \left(\chi\Ric[g]+\frac14K^2\right).
 \label{eq:hs-scalar-ct}
\end{equation}
Indeed, the Weyl variations are
\begin{align}
 \delta_\sigma\int_\Sigma\sqrt\gamma\,\Ric[g]
 &=-4\int_\Sigma\sqrt\gamma\,
   \left(\sigma_{nn}+K\sigma_n+\Delta_\Sigma\sigma\right),
 \notag\\
 \delta_\sigma\int_\Sigma\sqrt\gamma\,K^2
 &=4\int_\Sigma\sqrt\gamma\,K\sigma_n,
\end{align}
where $\Delta_\Sigma$ is the intrinsic Laplacian.
Taking Euclidean time to be periodic before sending
$\beta\to\infty$ makes $\Sigma$ closed, so the integrated
tangential Laplacian vanishes.
The counterterm variation then cancels
Eq.~\eqref{eq:hs-scalar-trivial}.
Both coefficients are thus fixed by the local anomaly,
independently of the vacuum-energy relation.
The $K^2$ term is required for this general matching even
though it contributes no energy on the hemisphere cylinder.

On this background, $K_{ab}=0$, $\Ric[g]=2/R^2$, and
$\int_\Sigma\dd^2y\sqrt\gamma=2\pi R\beta$.
Consequently, $W_\chi^{\rm ct}/\beta=-\chi/(64R)$, and
\begin{equation}
 E_\chi^{\rm matched}
 =\frac{\chi}{96R}-\frac{\chi}{64R}
 =-\frac{\chi}{192R}
 =-\frac{b_\chi}{12R},
 \label{eq:hs-scalar-matched-energy}
\end{equation}
where $b_\chi=\chi/16$.



\begin{thebibliography}{99}
\bibitem{Blote}
H.~W.~J. Bl\"ote, J.~L. Cardy, and M.~P. Nightingale,
Conformal invariance, the central charge, and universal finite-size
amplitudes at criticality,
\href{https://doi.org/10.1103/PhysRevLett.56.742}{Phys. Rev. Lett. \textbf{56}, 742 (1986)}.

\bibitem{Affleck}
I. Affleck, Universal term in the free energy at a critical point and
the conformal anomaly,
\href{https://doi.org/10.1103/PhysRevLett.56.746}{Phys. Rev. Lett. \textbf{56}, 746 (1986)}.

\bibitem{Billo}
M. Bill\`o, V. Gon\c{c}alves, E. Lauria, and M. Meineri,
Defects in conformal field theory,
\href{https://doi.org/10.1007/JHEP04(2016)091}{J. High Energy Phys. \textbf{04} (2016) 091},
\arxiv{1601.02883}.

\bibitem{JensenEntropy}
K. Jensen, A. O'Bannon, B. Robinson, and R. Rodgers,
From the Weyl anomaly to entropy of two-dimensional boundaries and defects,
\href{https://doi.org/10.1103/PhysRevLett.122.241602}{Phys. Rev. Lett. \textbf{122}, 241602 (2019)},
\arxiv{1812.08745}.

\bibitem{GrahamWitten}
C.~R. Graham and E. Witten, Conformal anomaly of submanifold
observables in AdS/CFT correspondence,
\href{https://doi.org/10.1016/S0550-3213(99)00055-3}{Nucl. Phys. B \textbf{546}, 52 (1999)},
\arxiv{hep-th/9901021}.

\bibitem{Henningson}
M. Henningson and K. Skenderis, Weyl anomaly for Wilson surfaces,
\href{https://doi.org/10.1088/1126-6708/1999/06/012}{J. High Energy Phys. \textbf{06} (1999) 012},
\arxiv{hep-th/9905163}.

\bibitem{Schwimmer}
A. Schwimmer and S. Theisen, Entanglement entropy, trace anomalies
and holography,
\href{https://doi.org/10.1016/j.nuclphysb.2008.04.015}{Nucl. Phys. B \textbf{801}, 1 (2008)},
\arxiv{0802.1017}.

\bibitem{Metlitski}
M.~A. Metlitski,
``Boundary criticality of the $O(N)$ model in $d=3$
critically revisited,''
SciPost Phys.\ \textbf{12}, 131 (2022)
[arXiv:2009.05119 [cond-mat.str-el]].

\bibitem{GiombiLiu}
S. Giombi and B. Liu, Notes on a surface defect in the $O(N)$ model,
\href{https://doi.org/10.1007/JHEP12(2023)004}{J. High Energy Phys. \textbf{12} (2023) 004},
\arxiv{2305.11402}.

\bibitem{Trepanier}
M. Tr\'epanier, Surface defects in the $O(N)$ model,
\href{https://doi.org/10.1007/JHEP09(2023)074}{J. High Energy Phys. \textbf{09} (2023) 074},
\arxiv{2305.10486}.

\bibitem{Raviv-Moshe:2023yvq}
A.~Raviv-Moshe and S.~Zhong,
``Phases of surface defects in Scalar Field Theories,''
JHEP \textbf{08}, 143 (2023)
doi:10.1007/JHEP08(2023)143
[arXiv:2305.11370 [hep-th]].

\bibitem{BianchiRenyi}
L. Bianchi, M. Meineri, R.~C. Myers, and M. Smolkin,
R\'enyi entropy and conformal defects,
\href{https://doi.org/10.1007/JHEP07(2016)076}{J. High Energy Phys. \textbf{07} (2016) 076},
\arxiv{1511.06713}.

\bibitem{DrukkerTech}
N. Drukker, M. Probst, and M. Tr\'epanier,
Defect CFT techniques in the 6d $\mathcal N=(2,0)$ theory,
\href{https://doi.org/10.1007/JHEP03(2021)261}{J. High Energy Phys. \textbf{03} (2021) 261},
\arxiv{2009.10732}.

\bibitem{HerzogHuang}
C.~P. Herzog and K.-W. Huang, Stress tensors from trace anomalies
in conformal field theories,
\href{https://doi.org/10.1103/PhysRevD.87.081901}{Phys. Rev. D \textbf{87}, 081901(R) (2013)},
\arxiv{1301.5002}.

\bibitem{Assel}
B. Assel, D. Cassani, L. Di Pietro, Z. Komargodski, J. Lorenzen,
and D. Martelli, The Casimir energy in curved space and its
supersymmetric counterpart,
\href{https://doi.org/10.1007/JHEP07(2015)043}{J. High Energy Phys. \textbf{07} (2015) 043},
\arxiv{1503.05537}.

\bibitem{Bobev}
N. Bobev, M. Bullimore, and H.-C. Kim,
Supersymmetric Casimir energy and the anomaly polynomial,
\href{https://doi.org/10.1007/JHEP09(2015)142}{J. High Energy Phys. \textbf{09} (2015) 142},
\arxiv{1507.08553}.

\bibitem{Huang}
Z.-X. Huang, M.-K. Yuan, and Y. Zhou,
From Weyl anomaly to universal defect Casimir energy and R\'enyi entropy,
\href{https://doi.org/10.1103/fz81-ysss}{Phys. Rev. Lett. \textbf{136}, 201601 (2026)},
\arxiv{2501.09498}.

\bibitem{DeserSchwimmer}
S. Deser and A. Schwimmer, Geometric classification of conformal
anomalies in arbitrary dimensions,
\href{https://doi.org/10.1016/0370-2693(93)90934-A}{Phys. Lett. B \textbf{309}, 279 (1993)},
\arxiv{hep-th/9302047}.

\bibitem{Asnin}
V. Asnin, Analyticity properties of Graham--Witten anomalies,
\href{https://doi.org/10.1088/0264-9381/25/14/145013}{Classical Quantum Gravity \textbf{25}, 145013 (2008)},
\arxiv{0801.1469}.

\bibitem{DrukkerNonlinear}
N. Drukker, Z. Kong, and P. Kravchuk,
Nonlinearly realised defect symmetries and anomalies,
\arxiv{2512.15913}.

\bibitem{Lanzetta:2025xfw}
R.~A.~Lanzetta, S.~Liu and M.~A.~Metlitski,
``The beginning of the endpoint bootstrap for conformal line defects,''
[arXiv:2508.14964 [cond-mat.str-el]].

\bibitem{HerzogHuangJensen}
C.~P. Herzog, K.-W. Huang, and K. Jensen,
Universal entanglement and boundary geometry in conformal field theory,
\href{https://doi.org/10.1007/JHEP01(2016)162}{J. High Energy Phys. \textbf{01} (2016) 162},
\arxiv{1510.00021}.

\bibitem{Gustavsson}
A. Gustavsson, On the Weyl anomaly of Wilson surfaces,
\href{https://doi.org/10.1088/1126-6708/2003/12/059}{J. High Energy Phys. \textbf{12} (2003) 059},
\arxiv{hep-th/0310037}.

\bibitem{DrukkerSurface}
N.~Drukker, M.~Probst, and M.~Tr\'epanier,
Surface operators in the 6d $\mathcal N=(2,0)$ theory,
J. Phys. A \textbf{53}, 365401 (2020),
arXiv:2003.12372.

\bibitem{KlebanovTseytlin1996}
I.~R. Klebanov and A.~A. Tseytlin,
``Entropy of near-extremal black $p$-branes,''
Nucl.\ Phys.\ B \textbf{475}, 164--178 (1996)
[arXiv:hep-th/9604089].

\bibitem{Tseytlin2000}
A.~A. Tseytlin,
``$R^4$ terms in 11 dimensions and conformal anomaly
of (2,0) theory,''
Nucl.\ Phys.\ B \textbf{584}, 233--250 (2000)
[arXiv:hep-th/0005072].

\bibitem{Diatlyk}
O. Diatlyk, Z. Sun, and Y. Wang,
Surprises in the ordinary: $O(N)$ invariant surface defect in the
$\epsilon$-expansion,
\href{https://doi.org/10.1007/JHEP06(2025)131}{J. High Energy Phys. \textbf{06} (2025) 131},
\arxiv{2411.16522}.

\bibitem{Monodromy}
L. Bianchi, A. Chalabi, V. Proch\'azka, B. Robinson, and J. Sisti,
Monodromy defects in free field theories,
\href{https://doi.org/10.1007/JHEP08(2021)013}{J. High Energy Phys. \textbf{08} (2021) 013},
\arxiv{2104.01220}.

\bibitem{Hyperbolic}
S. Giombi, E. Helfenberger, Z. Ji, and H. Khanchandani,
Monodromy defects from hyperbolic space,
\href{https://doi.org/10.1007/JHEP02(2022)041}{J. High Energy Phys. \textbf{02} (2022) 041},
\arxiv{2102.11815}.

\bibitem{LewkowyczPerlmutter}
A. Lewkowycz and E. Perlmutter,
Universality in the geometric dependence of R\'enyi entropy,
\href{https://doi.org/10.1007/JHEP01(2015)080}{J. High Energy Phys. \textbf{01} (2015) 080},
\arxiv{1407.8171}.

\bibitem{DrukkerTrepanier2022Crease}
N.~Drukker and M.~Tr\'epanier,
``Ironing out the crease,''
JHEP \textbf{08} (2022), 193,
doi:10.1007/JHEP08(2022)193
[arXiv:2204.12627 [hep-th]].

\bibitem{DrukkerTrepanier2021BPS}
N.~Drukker and M.~Tr\'epanier,
``Observations on BPS observables in 6d,''
J.\ Phys.\ A \textbf{54} (2021), 205401,
doi:10.1088/1751-8121/abf38d
[arXiv:2012.11087 [hep-th]].

\bibitem{DrukkerForini}
N.~Drukker and V.~Forini,
``Generalized quark-antiquark potential at weak and strong coupling,''
JHEP \textbf{06} (2011), 131,
doi:10.1007/JHEP06(2011)131
[arXiv:1105.5144 [hep-th]].

\bibitem{Cuomo:2024psk}
G.~Cuomo, Y.~C.~He and Z.~Komargodski,
``Impurities with a cusp: general theory and 3d Ising,''
JHEP \textbf{11}, 061 (2024)
doi:10.1007/JHEP11(2024)061
[arXiv:2406.10186 [hep-th]].

\bibitem{JensenRG}
K. Jensen and A. O'Bannon,
Constraint on defect and boundary renormalization group flows,
\href{https://doi.org/10.1103/PhysRevLett.116.091601}{Phys. Rev. Lett. \textbf{116}, 091601 (2016)},
\arxiv{1509.02160}.

\bibitem{KrishnanMetlitski}
A.~Krishnan and M.~A.~Metlitski,
A plane defect in the 3d $O(N)$ model,
\href{https://doi.org/10.21468/SciPostPhys.15.3.090}
{SciPost Phys. \textbf{15}, 090 (2023)},
\arxiv{2301.05728}.

\bibitem{Georgiou}
G. Georgiou, Weyl anomaly coefficients of holographic defect CFTs
at weak and strong coupling,
\href{https://doi.org/10.1007/JHEP08(2026)138}{J. High Energy Phys. \textbf{08} (2026) 138},
\arxiv{2604.19881}.

\bibitem{FuzzySphere}
W. Zhu, C. Han, E. Huffman, J. S. Hofmann, and Y.-C. He,
Uncovering conformal symmetry in the 3D Ising transition:
State-operator correspondence from a fuzzy sphere regularization,
\href{https://doi.org/10.1103/PhysRevX.13.021009}{Phys. Rev. X \textbf{13}, 021009 (2023)},
\arxiv{2210.13482}.

\bibitem{FuzzySurface}
Z. Zhou and Y. Zou,
Studying the 3d Ising surface CFTs on the fuzzy sphere,
\href{https://doi.org/10.21468/SciPostPhys.18.1.031}{SciPost Phys. \textbf{18}, 031 (2025)},
\arxiv{2407.15914}.

\bibitem{Feng:2026iii}
J.~Feng and T.~Wang,
``Studying 3D O(N) Surface CFT on the Fuzzy Sphere,''
[arXiv:2604.21091 [cond-mat.str-el]].

\bibitem{Dedushenko:2024nwi}
M.~Dedushenko,
``Ising BCFT from Fuzzy Hemisphere,''
[arXiv:2407.15948 [hep-th]].

\bibitem{RydbergIsing}
S. Ebadi \textit{et al.},
Quantum phases of matter on a 256-atom programmable quantum
simulator,
\href{https://doi.org/10.1038/s41586-021-03582-4}{Nature \textbf{595}, 227--232 (2021)},
\arxiv{2012.12281}.

\bibitem{RydbergLocal}
T. Manovitz \textit{et al.},
Quantum coarsening and collective dynamics on a programmable
simulator,
\href{https://doi.org/10.1038/s41586-024-08353-5}{Nature \textbf{638}, 86 (2025)},
\arxiv{2407.03249}.

\bibitem{Casini}
H. Casini, M. Huerta, and R.~C. Myers,
Towards a derivation of holographic entanglement entropy,
\href{https://doi.org/10.1007/JHEP05(2011)036}{J. High Energy Phys. \textbf{05} (2011) 036},
\arxiv{1102.0440}.

\bibitem{JensenHolography}
K. Jensen and A. O'Bannon,
Holography, entanglement entropy, and conformal field theories with
boundaries or defects,
\href{https://doi.org/10.1103/PhysRevD.88.106006}{Phys. Rev. D \textbf{88}, 106006 (2013)},
\arxiv{1309.4523}.

\bibitem{Kobayashi:2018lil}
N.~Kobayashi, T.~Nishioka, Y.~Sato and K.~Watanabe,
``Towards a $C$-theorem in defect CFT,''
JHEP \textbf{01}, 039 (2019)
doi:10.1007/JHEP01(2019)039
[arXiv:1810.06995 [hep-th]].

\bibitem{Chalabi:2020iie}
A.~Chalabi, A.~O'Bannon, B.~Robinson and J.~Sisti,
``Central charges of 2d superconformal defects,''
JHEP \textbf{05}, 095 (2020)
doi:10.1007/JHEP05(2020)095
[arXiv:2003.02857 [hep-th]].

\bibitem{Wang}
Y. Wang, Surface defect, anomalies and $b$-extremization,
\href{https://doi.org/10.1007/JHEP11(2021)122}{J. High Energy Phys. \textbf{11} (2021) 122},
\arxiv{2012.06574}.

\bibitem{HerzogShamir}
C.~P. Herzog and I. Shamir, Anomalies from correlation functions
in defect conformal field theory,
\href{https://doi.org/10.1007/JHEP07(2021)091}{J. High Energy Phys. \textbf{07} (2021) 091},
\arxiv{2103.06311}.

\bibitem{SimmonsDuffin:2016gjk}
D.~Simmons-Duffin,
\textit{TASI Lectures on the Conformal Bootstrap},
arXiv:1602.07982 [hep-th].

\bibitem{VassilevichHeatKernel}
D.~V. Vassilevich, Heat kernel expansion: User's manual,
Phys. Rep. \textbf{388}, 279--360 (2003),
doi:10.1016/j.physrep.2003.09.002, arXiv:hep-th/0306138.

\end{thebibliography}
\end{document}